\documentclass[12pt,a4paper]{article}

\usepackage{graphicx}

\newcommand{\be}{\begin{equation}}

\newcommand{\ee}{\end{equation}}
\newcommand{\ba}{\begin{eqnarray}}
\newcommand{\ea}{\end{eqnarray}}

\newcommand{\e}{{\rm e}}
\newcommand{\ie}{{\em i.e.\ }}

\newcommand{\tr}{{\rm tr}}
\newcommand{\bD}{\bar{D}}

\newcommand{\cN}{{\cal N}}
\newcommand{\cK}{{\cal K}}
\newcommand{\cO}{{\cal O}}
\newcommand{\cL}{{\cal L}}

\newcommand{\cV}{{\cal V}}

\newcommand{\cI}{{\cal I}}
\newcommand{\cW}{{\cal W}}

\renewcommand{\P}{\Phi}

\newcommand{\dP}{\Phi^{\dagger}}

\renewcommand{\a}{\alpha}

\renewcommand{\b}{\beta}

\renewcommand{\d}{\delta}
\newcommand{\q}{\theta}
\newcommand{\bq}{\bar\q}

\newcommand{\ep}{\epsilon}

\newcommand{\bt}[1]{{\bar t}}

\newcommand{\lan}{\langle}
\newcommand{\ran}{\rangle}
\newcommand{\pa}{\partial}
\newcommand{\g}{\gamma}
\renewcommand{\o}{\omega}
\begin{document}
\begin{titlepage}
\date{}
\vspace{-2.0cm}
\title{
{\vspace*{-6mm} 
\begin{flushright}
\small{LAPTH-1153 }\\
\small{ROM2F/2006/13}
\end{flushright}
\vspace{10mm}
}
On the all-order perturbative finiteness of the deformed ${\cal N}=4$ SYM theory
 \vspace*{0mm}}
\author{G.C.\ Rossi$^{a}$,\quad E.\ Sokatchev$^{b}$,
\quad Ya.S.\ Stanev$^{a}$ \\[5mm]
  {\small $^a$ Dipartimento di Fisica, Universit\`a di  Roma
   ``{\it Tor Vergata}''} \\
  {\small and INFN, Sezione di Roma ``{\it Tor Vergata}''}\\
  {\small Via della Ricerca Scientifica - 00133 Roma, Italy}\\
  {\small $^b$  Laboratoire d'Annecy-le-Vieux de
Physique Th\'{e}orique  LAPTH,}\\
  {\small B.P. 110,  F-74941 Annecy-le-Vieux,
  France\footnote{UMR 5108 associ\'{e}e \`{a}
 l'Universit\'{e} de Savoie 
 } }}
\maketitle
\vspace*{0mm}

\abstract {We prove that the chiral propagator of the deformed ${\cal N}=4$ SYM theory
can be made finite to all orders in perturbation theory for any complex value of the deformation
parameter.  For any such value
the set of finite deformed theories can be parametrized
by a whole complex function of the coupling constant $g$.
We reveal a new protection mechanism for chiral operators of dimension three.
These are obtained by differentiating the Lagrangian with respect to the
independent coupling constants. A particular combination of them is a CPO
involving only chiral matter. Its all-order form is derived directly
from the finiteness condition. The procedure is confirmed perturbatively through order $g^6$.}

\vspace{1truecm}
\vspace{2mm} \vfill \hrule width 3.cm
\begin{flushleft}
e-mail: giancarlo.rossi@roma2.infn.it\\
e-mail: emeri.sokatchev@cern.ch\\
e-mail: yassen.stanev@roma2.infn.it
\end{flushleft}

\end{titlepage}

\vfill
\newpage

\section {Introduction}

Recently, following~\cite{LS,BJL,LuninM},  there has been a renewed
interest in the deformed $\cN=4$
supersymmetric Yang-Mills (SYM) theory. It has the same field content as $\cN=4$ SYM,
namely (in an $\cN=1$ formulation) a gauge superfield $V$ and a set of three chiral
matter superfields $\Phi^{I}$, $I=1,2,3$, all in the adjoint representation of the
gauge group  $SU(N)$. What distinguishes the deformed theory from $\cN=4$ SYM is the deformed
superpotential
\be
{\cal W} =  g \,  \kappa \ \tr \left(
\Phi^{1} \ [\Phi^{2}, \Phi^{3}]_{\omega} \right)
\,   ,
\label{old_superp}
\ee
where $g$ is the $\cN=1$ SYM coupling constant and the
deformed commutator is defined as
\be
  [A,B]_{\omega} = \omega \ AB -  BA \, .
\label{defcom}
\ee
The parameter $\kappa$ can be considered real (its phase can be absorbed into a
redefinition of $\P^I$), while $\omega$
is in general complex. The undeformed $\cN=4$ SYM is
recovered when $\kappa =\omega= 1$.
Although in principle both $\kappa$ and $\omega$ can be taken as 
Taylor series expansions in powers of $g^2$ around $g=0$, in
most of the recent literature~\cite{FG,PSZ,Nie,PSZ2,PSZ3} the case of constant
$\omega$ has been commonly considered.

The main feature of the deformed theory is that, despite the breaking of $\cN=4$
supersymmetry down to $\cN=1$, it can be made finite (and thus conformal) by imposing
a condition on the parameters $g,\kappa,\o$. The search for finite $\cN=1$ theories
has a long history~\cite{history,Parkes,Ermushev:1986cu,Kazakov:1986vg,JackJN}.
In the most general case one considers a superpotential of the  Yukawa type
\begin{equation}\label{gensupepo}
    Y_{ijk} \P^i\P^j\P^k\,,
\end{equation}
where $i,j,k$ are combined color and flavor indices, and $Y_{ijk}$ is a set of complex couplings.
These theories are finite if all beta and gamma functions vanish. In the matter
sector $\b_Y$ and $\g_Y$ are related through the non-renormalization of the chiral
vertex~\cite{nonchi,GRS}, so it is sufficient to demand the vanishing of the
matrix of gamma functions of the chiral superfields, $\g_\P(g,Y)^i_j=0$.
This is a set of conditions on the couplings which are to be adjusted
order by order in perturbation theory. The existence of a solution
in the general case has been studied in~\cite{Ermushev:1986cu,Kazakov:1986vg}.

The superpotential~(\ref{old_superp}) is a particular case of~(\ref{gensupepo})
with the interesting feature that all matter gamma functions are equal due to
the $Z_3\times Z_3$ symmetry
of the potential\footnote{The most general superpotential
with this property was found in~\cite{LS,Aharony:2002hx}; see Section \ref{genpotsing}
for more details.}. So, it is enough to impose a {\it single finiteness condition}
\begin{equation}\label{singfi}
    \g_\P(g,\kappa,\o) = 0
\end{equation}
to ensure that the matter propagators and couplings do not undergo infinite renormalization.
This feature of the so-called ``$\b$--deformed $\cN=4$ theory" with superpotential~(\ref{old_superp})
was essential for finding its gravity dual in~\cite{LuninM} and for the subsequent development
in the context of the AdS/CFT correspondence~\cite{subseq,Khoze}.

The question about the vanishing of the propagator corrections and of the beta
function in the $\cN=1$ gauge sector of theories with the superpotential~(\ref{gensupepo})
is more subtle. A three-loop result is available~\cite{Parkes}, but its generalization to
all orders~\cite{JackJN,LS} relies on the existence of the so-called ``exact $\b$
function"~\cite{Novikov:1983uc}.

The first steps in the study of the perturbative aspects of the deformed theory
with superpotential~(\ref{old_superp}) were made in~\cite{FG}, with a particular
accent on the chiral primary operators (CPO) in the theory. Subsequently, the
condition for finiteness were established at two loops in~\cite{PSZ} and at
three loops in~\cite{Nie}. An all-order condition in the large $N$ limit was
found in~\cite{PSZ2}. The set of CPOs (``chiral ring") of the deformed theory
was further studied in~\cite{PSZ,Nie,PSZ2,PSZ3}.

In the present paper we concentrate on two particular perturbative issues
in the deformed  $\cN=4$ SYM theory.

In Section \ref{sec:Allloops} we investigate the finiteness properties
of a theory with superpotential
${\cal W}$, deformed by a $g$-dependent deformation parameter.
For future convenience we shall write it in the
form\footnote{The trace is over the color indices of
the fundamental representation of the $SU(N)$ gauge group. The
generators, $T^{a}$, of the fundamental representation are
normalized according to $\tr(T^{a}T^{b})=\frac{1}{2}
\delta^{ab}$.}
\ba
{\cal W_{\kappa , \, \omega}} & = & g  \left\{
\kappa_{\omega}(g) \ \tr \left( \Phi^{1} \ [\Phi^{2},
\Phi^{3}]_{\omega} \right) \ + \ \kappa_{\Omega}(g) \ \tr \left(
\Phi^{1} \ [\Phi^{2},
\Phi^{3}]_{\Omega}  \right) \ \right\} \nonumber \\
& + & g  \left\{ \kappa_{ \omega}(g) \  \tr \left(
\Phi_{1}^{\dagger} \ [\Phi_{3}^{\dagger},
\Phi_{2}^{\dagger}]_{\bar \omega} \right) \ + \ \bar
\kappa_{\Omega}(g) \ \tr \left( \Phi_{1}^{\dagger} \
[\Phi_{3}^{\dagger}, \Phi_{2}^{\dagger}]_{\bar \Omega}  \right) \
\right\} \, ,
\label{superp}
\ea
where $\omega$ is a complex constant\footnote{The
$g$-dependence of the deformation parameter is
hidden in the terms proportional to $[ \ , \ ]_{\Omega}$.
Indeed one can rewrite~(\ref{superp}) in the form of~(\ref{old_superp})
with a $g$-dependent deformation parameter $\omega(g)$.},
$\bar \omega$ is its complex
conjugate and  $\Omega$  and $\bar \Omega$ are defined as
\be
\Omega = - { N^2-2 +2 \ \bar \omega \over (N^2-2)\ \bar \omega +2}
\quad ,  \qquad \bar \Omega = - { N^2-2 +2 \  \omega \over
(N^2-2)\ \omega +2} \, .
\label{Omega}
\ee
The main  result of Section \ref{sec:Allloops} is the proof that for {\it any} complex
constant $\omega$ and {\it any} complex function
$\kappa_{\Omega}(g)$ satisfying $\kappa_{\Omega}(0) = 0 $, there
exists  a {\it unique} function $\kappa_{\omega}(g)$, such that the
chiral propagator is finite to all orders in perturbation theory,
with the consequence that
 the chiral field has a vanishing anomalous dimension.
To be precise, since we shall
always compute the difference between the quantities in the
deformed theory and in ${\cal N} = 4$ SYM (which corresponds to 
$\kappa_{\omega}=\omega=1$, $\kappa_{\Omega}=0$), everywhere in
this paper by ``finite"  we actually mean ``as finite as in ${\cal N}
= 4$ SYM". We derive the explicit form of the finiteness condition at
order $g^6$ for any number of colors $N$,
and at order $g^8$ in the planar $N \rightarrow \infty$ limit.
We also briefly discuss the corrections to the three-point vertices.

In Section \ref{Free} we study a particular type of CPO of dimension three, namely
\begin{equation}\label{defFr}
    {\cO_F} = \tr \left(\Phi^{1}\Phi^{2} \Phi^{3} \right) + \a\, \tr \left(
\Phi^{1}\Phi^{3} \Phi^{2} \right)\,,
\end{equation}
which is a mixture of the two terms in the superpotential. Its existence was
revealed in~\cite{FG} where the one-loop value of the mixing coefficient $\a$ was
determined through a direct two-point function computation. This one-loop result
was confirmed in~\cite{PSZ,Nie} and in~\cite{PSZ3} it was shown that $\a$ is not corrected
at two loops. We compute the  three-loop correction to the value of $\a$.
However, the main purpose of Section \ref{Free} is to show that
$\a$ can in fact be determined {\it without graph calculations}, but directly
from the finiteness condition~(\ref{singfi}). The key observation is that the
quantum corrections to the correlators of composite operators, i.e. their
derivatives with respect to each independent coupling, are generated by the
insertion of very special CPOs of the type $\cI = a\, \tr(W^2) + b\, \tr(\P\P\P)$
(here $\tr(W^2)$ is the $\cN=1$ SYM chiral Lagrangian). The latter are obtained
by differentiation of the chiral part of the Lagrangian, taking into account the
relation among the couplings. To do this we rewrite the superpotential in
the form $\cW = f\, \tr \left(\Phi^{1}[\Phi^{2}, \Phi^{3}] \right) + d\,
\tr\left(\Phi^{1}{\{}\Phi^{2}, \Phi^{3}{\}} \right)$ and treat the holomorphic
couplings $f,d$ as independent, while $g$ is determined from the finiteness
condition $\g_\P(g,f,d,\bar f, \bar d) =0$. The derivatives with respect to
$f,d$ give rise to two CPOs $\cI_{f,d}$. We can say that $\cI_{f,d}$ generate
quantum corrections along the tangent directions to the surface in the moduli
space defined by $\g_\P=0$.  Then the operator $\cO_F$ is simply the linear
combination of $\cI_f$ and $\cI_d$ such that $\tr(W^2)$ drops out. This means that
the form of $\cO_F$ to {\it all orders} is directly determined by the corresponding
finiteness condition $\g_\P=0$. When restricted to three loops, the general formula exactly
reproduces the result of our graph calculation. Also, we can
immediately explain the observation  of~\cite{PSZ3}
that $\a$ is not corrected
at two loops - it simply follows from the fact that $\g_\P$ has no two-loop
contribution.  In Section \ref{genpotsing} we generalize the construction of $\cO_F$ to the most 
general deformed theory which is made finite by a single condition~\cite{LS,Aharony:2002hx}.

\section{All-order perturbative finiteness }
\label{sec:Allloops}


Before proceeding, let us briefly motivate our conventions. The two
deformed commutators $[ \ , \ ]_{\omega}$ and $[ \ , \ ]_{\Omega}$
in eq.~(\ref{superp}) are just a conventional choice of basis in
the two dimensional space of color structures (alternative to
$f_{abc}$ and $d_{abc}$, which correspond to the choices $\omega=1$ and
$\Omega=-1$, respectively). The explicit form of $\Omega$ given in
eq.~(\ref{Omega}) is determined by the requirement that $[ \ , \
]_{\Omega}$ and $[ \ , \ ]_{\bar \omega}$  are orthogonal in the
sense that
\be
\sum_{c,d} \ \tr ( T^a \ [ T^c , T^d ]_{\Omega}) \ \tr( T^b \ [
T^d , T^c ]_{\bar \omega}) \  = \ 0 \, ,
\label{Oort}
\ee
and similarly for $\bar \Omega$ and $\omega$. This implies also the
vanishing of the color contractions of the $\omega$ and $\bar \Omega$
(as well as the $\bar \omega$ and $\Omega$)  terms in the
superpotential ${\cal W_{\kappa , \, \omega}}$ of eq.~(\ref{superp}). Note also that if
$\omega$ is a pure phase $| \, \omega | =1 $, then also $| \,
\Omega | =1 $.
 The real
function $\kappa_{\omega}(g)$ and the complex function
$\kappa_{\Omega}(g)$ have a power series expansion in $g^2$
that we find useful to cast in the
form
\ba
\kappa_{\omega}(g) & = & \sum_{n=0}^{\infty} \kappa_{\omega}^{(n)}
(g^{2} N)^{n} \, , \\
&  &  \nonumber \\
 \kappa_{\Omega}(g) & = & \sum_{n=1}^{\infty}
\kappa_{\Omega}^{(n)} (g^{2} N)^{n} \ .
\label{KoO}
\ea
Note that by assumption $\kappa_{\Omega}(0)=0$, while
$\kappa_{\omega}(0)=\kappa_{\omega}^{(0)} \neq 0$.
At each order in $g^2$ the general superpotential ${\cal W_{\kappa , \, \omega}}$
depends on 3 real parameters. For
 $n>0$ they are $\kappa_{\omega}^{(n)}$ (real)  and
$\kappa_{\Omega}^{(n)}$ (complex), while for $n=0$ we choose them
to be $\kappa_{\omega}^{(0)}$ (real)  and $\omega$ (complex).
As
we shall see,  this choice allows us to express in a compact form the
solution of the condition for the perturbative finiteness  of the chiral field propagator to all
orders.

\subsection{The chiral propagator to all orders}
\label{sec:Allorders}

We start by reconsidering the  order $g^2$, $g^4$ and $g^6$
conditions for finiteness of the chiral propagator in the
theory with the general
 superpotential ${\cal W_{\kappa , \, \omega}}$ of
 eq.~(\ref{superp}) (the order $g^6$ condition in the case of the
superpotential of  eq.~(\ref{old_superp}) was found
in~\cite{Nie}).

Let us  write the action of the deformed theory in the form
\be
S_{\kappa , \, \omega} = S_0 + S_v + S_{\cal W_{\kappa , \, \omega}} \ ,
\label{Sh}
\ee
where $S_0$ contains the free kinetic terms
 and $S_v$ is the part of
the standard $\cN=4$ SYM action involving the couplings of the
gauge superfield $V$ (including the gauge-fixing term).
Finally, $S_{\cal W_{\kappa , \, \omega}}$
is the part of the action involving the deformed superpotential
${\cal W_{\kappa , \, \omega}}$ given in
 eq.~(\ref{superp}).
In this notation the action of the undeformed $\cN=4$
SYM theory reads ($\kappa_{\omega}=\omega=1$, $\kappa_{\Omega}=0$)
\be
S_g = S_0 + S_v + S_{\cW_g} \, .
\label{Sg}
\ee
In the deformed theory the lowest $\theta$ components of the order $g^{2n}$
correction to the propagator of the chiral
superfield\footnote{Because of the $Z_3$ symmetry of the action,
all three chiral superfields are on the same footing, so our
choice of the flavor index is conventional.} $ \langle \Phi^1_{a}(x_1,
\theta_1) \Phi_{1 \ b}^{\dagger}(x_2,\bar\theta_2) \rangle$, can be compactly written
in the form
\be
G_{\kappa , \, \omega}^{2n}(x_1,x_2) =\langle\left. \e^{ S_v+
S_{\cal W_{\kappa , \, \omega}} }\right|_{g^{2n}} \phi^1_{a}(x_1) \phi_{1
\ b}^{\dagger}(x_2) \rangle \, ,
\label{Ghn}
\ee
where by $\e^{S_v+S_{\cal W_{\kappa , \, \omega}}}|_{g^{2n}}$ we denote
all terms of order $g^{2n}$ in the expansion of the exponent. The
similar computation, which in $\cN=4$ SYM reads
\be
G_g^{2n} (x_1, x_2) =\langle\left. \e^{S_v+S_{\cW_g}}\right|_{g^{2n}}
\phi^1_{a}(x_1) \phi_{1 \ b}^{\dagger}(x_2)\rangle \, ,
\label{Ggn}
\ee
 is know to give  a  finite result~\cite{FiniteN4}. Hence, if the computation
of the difference
\be
\delta G^{2n} (x_1,x_2) = G_{\kappa , \, \omega}^{2n} (x_1,x_2) -
G_g^{2n} (x_1,x_2) \, ,
\label{deltaG_1}
\ee
gives a finite result, then also $G_{\kappa , \,
\omega}^{2n}(x_1,x_2)$ will be finite. Note that computing
the difference is much simpler than each term separately,
since most of the vector field
contributions cancel out. In particular, as far as the chiral
propagator is concerned, up to  order $g^6$, effectively only
the superpotential contributes to the difference (for the details
see~\cite{Nie}), leaving the quantities
\be
\delta G^{2n} (x_1,x_2) = \langle
 \left(\left. e^{S_{ \cW_{\kappa , \, \omega} }}\right|_{g^{2n}}
 - \left. e^{S_{ \cW_g }} \right|_{g^{2n}} \right)
\phi^1_{a}(x_1) \phi_{1 \ b}^{\dagger}(x_2)\rangle \,  ,
\label{deltaG_2}
\ee
 to be evaluated for $n=1,2,3$.

Moreover, at each perturbative order  we just have to
take into account  the
primitive divergent superdiagrams (\ie those which do not contain
divergent subdiagrams).
\begin{figure} [ht]
    \centering
    \includegraphics[width=1\linewidth]{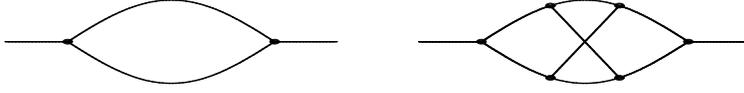}
       \vskip -220pt
\caption{The one-loop and the three-loop diagrams.}
  \label{diagr1}
\end{figure}
The only two such contributions to~(\ref{deltaG_2}) up to order
$g^6$
are shown in Figure~1 and they are both logarithmically divergent~\cite{AG}. 
The first has
the topology of the one-loop diagram. The second is a genuine
three-loop (nonplanar) diagram. It is present only in the
deformed theory, because the corresponding color factor in
$\cN=4$ SYM is zero. Owing to the chiral structure of the superpotential
there are no primitive  divergent two-loop superdiagrams.

As we said,  there is a single divergent
superdiagram at order $g^2$. Thus finding the  finiteness condition at this order
reduces to a simple color contraction problem. The relevant color contractions are
\ba
\sum_{c,d} \ \tr ( T^a \ [ T^c , T^d ]_{\omega})  \  \tr( T^b \ [
T^d , T^c ]_{\bar \omega}) \ & = & \delta^{ab} \ C_{\omega}  \,
, \label{conoo} \\
\sum_{c,d} \ \tr ( T^a \ [ T^c , T^d ]_{\Omega})  \  \tr( T^b \ [
T^d , T^c ]_{\bar \omega}) \  & = & \ 0 \,
, \label{conoO} \\
\sum_{c,d} \ \tr ( T^a \ [ T^c , T^d ]_{\omega})  \  \tr( T^b \ [
T^d , T^c ]_{\bar \Omega}) \  & = & \ 0 \,
, \label{conOo} \\
\sum_{c,d} \ \tr ( T^a \ [ T^c , T^d ]_{\Omega})  \  \tr( T^b \ [
T^d , T^c ]_{\bar \Omega}) \ & = & \ \delta^{ab} \ C_{\Omega}  \,
.
\label{conOO}
\ea
where
\ba
C_{\omega}  & = &  \ {(N^2-2)(\omega \bar
\omega+1)+2(\omega+\bar \omega) \over 8 N } \, , \nonumber \\
 \label{CoO} \\
 C_{\Omega}  & =&   C_{\omega}
{N^2 (N^2-4) \over ((N^2-2)\omega +2)((N^2-2)\bar \omega +2)} \,
.\nonumber
\ea
The order $g^2$ (one-loop) finiteness condition then becomes
\be
(\kappa_{\omega}^{(0)})^2  \ C_{\omega}  - {N \over 4} \ = 0  \, ,
\label{oneloopv1}
\ee
or equivalently~\cite{FG,PSZ,Nie} (since $C_{\omega} > 0$ for any complex $\omega$
and integer $N>2$)
\be
(\kappa_{\omega}^{(0)})^2 \ = \ {2 \  N^2 \over
(N^2-2)(\omega\bar\omega+1)+2(\omega+\bar\omega)}
  \, ,
\label{1Lcond}
\ee
exactly as for the simpler case of the potential in
eq.~(\ref{old_superp}). Let us note that since the cancellation is
due to the vanishing of the numerical factor in front of one single
diagram, both the logarithmically divergent and  the finite part
contributions vanish.

At the next perturbative order  the difference
between the superpotentials of eqs.~(\ref{superp}) and
~(\ref{old_superp}) shows up. Indeed, if we impose
the constraint~(\ref{1Lcond}) (we always choose  the positive square
root for the solution) the order $g^4$ finiteness
condition reads
\be
\kappa_{\omega}^{(0)} \ \kappa_{\omega}^{(1)} \ C_{\omega} \ = \ 0
\, ,
\label{twoloopv1}
\ee
implying
\be
\kappa_{\omega}^{(1)} \ = 0
  \, .
\label{2Lcond}
\ee
Note that due to the vanishing of the contractions~(\ref{conoO}) and
~(\ref{conOo}) the complex coefficient $\kappa_{\Omega}^{(1)}$
remains undetermined. Hence, contrary to what is said in previous papers
the superpotential ${\cal W_{\kappa , \, \omega}}$
is allowed to  contain a different from zero
order $g^3$ term. Since the cancellation is again due to the
vanishing of numerical factors in front of each diagram,
both the divergent and the finite parts are set to zero
by eq.~(\ref{2Lcond}).

The situation becomes more complicated at order $g^6$,
since at this order both the genuine three-loop diagram and the one-loop
diagram (multiplied by order $g^4$ coefficients) contribute. Thus,
if we impose eq.~(\ref{2Lcond}),  the order $g^6$ finiteness
condition reads
\ba
2 \ \kappa_{\omega}^{(0)} \ \kappa_{\omega}^{(2)} \ C_{\omega} & +
&
 |\kappa_{\Omega}^{(1)}|^2  \ C_{\Omega} \  \label{3loopv1} \\
 && \nonumber \\
 & + &
 {3 \over 256 }  {\zeta(3) \ (k_{\omega}^{(0)})^6 \over  (4 \pi^2)^2}
 \ {(N^2-4)\over N^5}
 \ (\omega-1)\ (\bar \omega -1) \  P_3(\omega, N)  = \, 0 \, , \nonumber
\ea
where
\be
P_3(\omega, N) =  ( (\omega^2+\omega+1) ({\bar \omega}^2+\bar
\omega+1) - 9 \ \omega \bar \omega  ) \ N^2
 + 5 \ (\omega-1)^2 \ (\bar \omega-1)^2 \, ,
\label{P3}
\ee
and $C_{\omega}$ and $C_{\Omega}$ are given in eq.~(\ref{CoO}).  Note
that eq.~(\ref{3loopv1}) is linear in $\kappa_{\omega}^{(2)}$, so
there is always a unique solution for $\kappa_{\omega}^{(2)}$ as a
function of $\omega$, $N$ and $|\kappa_{\Omega}^{(1)}|$ for any
$\kappa_{\Omega}^{(1)}$. Let us stress that this is the first case
in which there are (nonvanishing) contributions from two different
(super)diagrams. In fact the first two terms in
eq.~(\ref{3loopv1}) come from the diagram with one-loop topology, while the third
term comes from the genuine  three-loop diagram.
Hence, the vanishing
of the divergent part does not automatically imply also the cancellation of the
(potentially scheme dependent) finite parts.
Let us stress that these finite contributions modify only the normalization of the
chiral superfield propagator.
This finite correction of the normalization, which is
present only in the deformed theory (remember that all terms in
eq.~(\ref{3loopv1}) originate from the deformed theory), will give rise
to a logarithmic divergency at the next order ($g^8$). Hence
starting at order $g^8$ the explicit form of the condition for
finiteness of the chiral propagator will in general be scheme
dependent.

It is clear that one can proceed iteratively. If we satisfy all
the finiteness conditions up to order $g^{2n}$, then  the
finiteness condition at order $g^{2(n+1)}$ will
schematically read
\be
2 \ \kappa_{\omega}^{(0)} \ \kappa_{\omega}^{(n+1)} \ C_{\omega} +
f_{n+1}(N,\omega,g,\{\kappa_{\Omega}^{(p)}, p=1,\dots,n \})
 \ = \ 0 \, ,
\label{AllLopps}
\ee
where $f_{n+1}$ is a computable  function and we have used the
lower order equations to express all $\kappa_{\omega}^{(q)}$,
$q=1,\dots,n$ in terms of the coefficients $\kappa_{\Omega}^{(p)}$,
$p=1,\dots,n-1$. Equation~(\ref{AllLopps}) is linear in
$\kappa_{\omega}^{(n+1)}$ and thus it has always a unique solution
(since both $ \kappa_{\omega}^{(0)}\neq 0 $ and $C_{\omega} \neq
0$).

To summarize we have shown that  for any complex constant $\omega$
and any complex function $\kappa_{\Omega}(g)$ satisfying
$\kappa_{\Omega}(0) = 0 $, there exist a {\it unique} (possibly
scheme dependent)  real function $\kappa_{\omega}(g)$, such that
the anomalous dimension of the chiral superfields is zero to all
orders in perturbation theory\footnote{Since the
potential (\ref{superp}) is a special case of  the general Yukawa
potential~(\ref{gensupepo}), our result could in principle be
obtained as a particular case of~\cite{Ermushev:1986cu}.
However, translating the parametrization from~(\ref{gensupepo})
to~(\ref{superp}) is not an easy task. Therefore we  find it
useful to give the explicit form of the finiteness
condition and its solution adapted to our special case.}.

The question of the finiteness of the vector superfield propagator
is more subtle and is beyond the scope of this paper.
It is closely related to the coupling constant
renormalization which we shall discuss in the next subsection.
We note that the finiteness of the vector superfield
propagator up to order $g^6$  follows only from the order $g^2$ and
$g^4$  conditions in eqs.~(\ref{1Lcond}) and~(\ref{2Lcond}).

\subsection{The three-point vertices}
\label{sec:3point}

Another important question which we would like to address is whether in
the deformed theory  there is a coupling
constant renormalization. To answer this question, one has to compute the
perturbative corrections to the three-point vertex functions.
There are four potentially different such vertex functions, namely the
triple chiral (or antichiral) vertex, the chiral-antichiral-vector
vertex, the triple vector vertex and the ghost-ghost-vector vertex
(see Figure~2).
\begin{figure} [ht]
    \centering
    \includegraphics[width=1\linewidth]{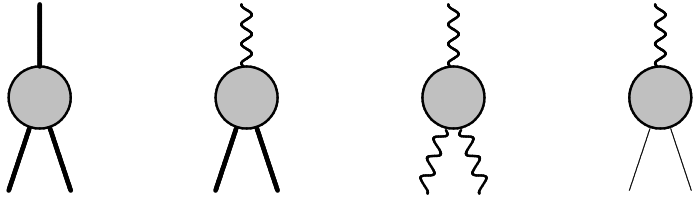}
     \vskip -180pt
\caption{The three-point vertices.}
  \label{three-point general }
\end{figure}
All four vertices in the action at tree level are proportional to $g$, but only the
triple chiral vertex at higher orders receives in addition to the
standard perturbative corrections also finite corrections from
$\kappa_{\omega}^{(n)}$ and $\kappa_{\Omega}^{(n)}$. Another
peculiar feature  of the triple chiral vertex is that its color
structure depends on $g$ in a non-trivial way. Indeed,
while the three vertices involving the vector field
are always proportional to the $SU(N)$ structure constants
$f^{abc}$, the triple chiral vertex is proportional to a linear
combination of the two deformed (by $\omega$ and $\Omega$)
commutators (see eq.~(\ref{superp})).

We recall that the triple chiral vertex function  obeys a non-renormalization theorem~\cite{nonchi,GRS}
which relates the propagator and the coupling constant renormalization factors.
The triple vector vertex is related to the simpler ghost-ghost-vector one by a
Slavnov-Taylor identity~\cite{TarasVladE280483}.
A similar identity relates the chiral-antichiral-vector vertex to
the matter propagator.

For our purposes  it again
suffices to compute only the difference between the values of  each  three-point
function  in the deformed
theory and in ${\cal N} =4$ SYM.  Our results can be summarized as
follows.

At order $g^3$  and order $g^5$ all
three  vertices with external vector lines are exactly equal to the
corresponding ones in ${\cal N} =4$ SYM. Only the triple chiral
vertex receives at order $g^5$ a finite non-planar correction
from the first (super)diagram in Figure 3. It
affects both the $[ \ , \ ]_{\omega}$ and $[ \ , \
]_{\Omega}$ structures in  the effective superpotential ${\cal W}_{\rm eff}$
\cite{PSZ3}.
In particular, the correction to the $[ \ , \ ]_{\omega}$  structure
is
\ba
{\cal W}_{\rm eff} |_{g^5\, , \, \omega}
&=& {3 \, \zeta(3) \ (k_{\omega}^{(0)})^5 \over 32 \, (4 \pi^2)^2}
 \ {(\omega-1)\ (\bar \omega -1)\
(N^2-4) \over N^2 \, ((N^2-2)(\omega \bar \omega+1)+2(\omega+\bar \omega))}
\nonumber \\
&& \nonumber \\
& \times & \ P_3(\omega, N) \
  \tr \left( \Phi^{1} \ [\Phi^{2}, \Phi^{3}]_{\omega}  \right) \, ,
\label{2L_omega}
\ea
with $P_3(\omega, N)$ defined in eq.~(\ref{P3}),
while the correction to
the $[ \ , \ ]_{\Omega}$ structure is
\ba
{\cal W}_{\rm eff} |_{g^5\, , \, \Omega}
&=& - \, {3 \, \zeta(3) \ (k_{\omega}^{(0)})^5 \over 32 \, (4 \pi^2)^2}
 \ {(\omega-1)( (N^2-2) \, \bar \omega
+2) \over N^2 \, ((N^2-2)(\omega \bar \omega+1)+2(\omega+\bar \omega))}
  \nonumber \\
&&  \nonumber \\
& \times & \ \left[ \, N^2 ((\omega+1) (\bar\omega^2-6\, \omega
\bar\omega+\omega^2) +4 \omega ( \omega \bar\omega^2+1) \right.
)   \label{2L_Omega}\\
&& \nonumber \\
& + &  \left. (\omega-1)^2 \, (\bar\omega-1) \, (7 (
\omega\bar\omega-1)-3\omega+3\bar\omega) \, \right] \,
  \tr \left( \Phi^{1} \ [\Phi^{2}, \Phi^{3}]_{\Omega}  \right)  . \nonumber
\ea
 At order $g^7$ vertices behave quite differently. The
ghost-ghost-vector vertex remains equal to its ${\cal N} =4$
value. The other three
vertices receive corrections from non-planar diagrams.
Whether they sum up to zero or not is an open
question\footnote{To answer it one has to carefully compute the
numerous three-loop super diagrams which contribute to the various
three-point functions. We shall address this issue in a future
publication.}. Only the triple chiral vertex receives also
finite planar corrections coming from the second (super)diagram
in Figure 3, which as explained in the next subsection modify
the chiral propagator at order $g^8$.

\begin{figure} [ht]
    \centering
    \includegraphics[width=1\linewidth]{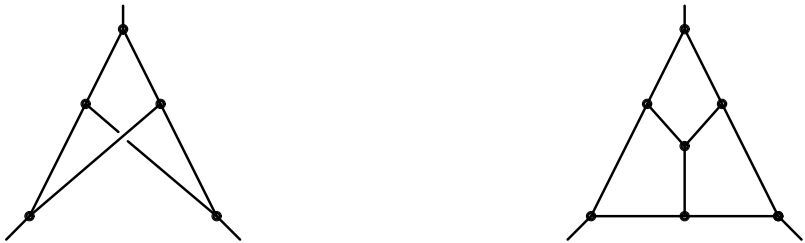}
     \vskip -180pt
\caption{The two-loop and the only  planar three-loop contributions to the triple
chiral vertex.}
  \label{three-point 3Loop}
\end{figure}

\subsection{The order $g^8$ condition in the planar limit}
\label{Sec:oderg^8}

In this subsection we shall derive the explicit form of the condition
for finiteness of the chiral superfield propagator at order $g^8$.
We shall work in the planar limit $N \rightarrow
\infty$, since this allows us to  drastically simplify the necessary
calculations. Still, the essential feature, namely the fact that
even in the planar limit, unless
$|\omega|=1$, one has to modify the coefficients in the superpotential
eq.~(\ref{superp}) by higher powers of $g$, clearly shows up.

Owing to the properties of the three-point vertices, in the planar
limit ($N \rightarrow \infty$ and $g^2 N$ fixed), there are significant
simplifications. On the one hand,  the order $g^8$
correction (with respect to the ${\cal N}=4$ SYM value) to the vector propagator is zero. Indeed all
diagrams which contribute will contain as a subdiagram some lower
order vertex with external vector lines which, as we mentioned in the previous subsection,
vanish in the planar limit. On the other hand the corrections to
the chiral propagator will come only from the diagram with the one
loop topology shown in Figure~1
and from the planar three-loop correction to the chiral vertex
shown in Figure~3, which leads to an order $g^8$
planar primitive logarithmically divergent diagram
shown in Figure~4.

\begin{figure} [ht]
    \centering
    \includegraphics[width=1\linewidth]{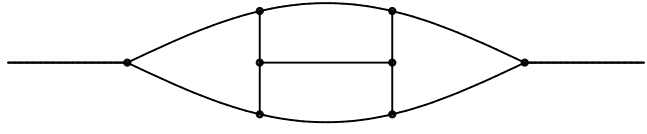}
       \vskip -220pt
\caption{The 4-loop primitive planar diagram.}
  \label{four loop }
\end{figure}

In the planar limit the finiteness conditions become
\ba
g^2 : &&(\kappa_{\omega}^{(0)})^2 \  =  \ {2 \over (1+\omega\bar\omega)}
  \ , \label{1L_planar} \\
  && \nonumber \\
g^4 : &&\kappa_{\omega}^{(1)} \ = \ 0 \ ,\label{2L_planar} \\
&& \nonumber \\
g^6 : && 2 \ \omega\bar\omega \ \kappa_{\omega}^{(0)} \
\kappa_{\omega}^{(2)} \   +
 |\kappa_{\Omega}^{(1)}|^2   \ = \ 0 \ ,
\label{3L_planar}
\ea
because $\Omega=-1/\bar\omega$,
$\bar\Omega=-1/\omega$, $C_{\omega}/C_{\Omega}=\omega\bar\omega$.
After imposing these conditions,
the cancellation of the order $g^8$ divergencies coming from the diagram
with one-loop topology and the planar 4-loop diagram depicted in
Figure 4, leads to the condition
\ba
 2  \ \kappa_{\omega}^{(0)} \
\kappa_{\omega}^{(3)} \   +
 {\kappa_{\Omega}^{(1)} \overline{\kappa_{\Omega}^{(2)}}+
 \kappa_{\Omega}^{(2)} \overline{\kappa_{\Omega}^{(1)}}
 \over \omega\bar\omega} \ & = &
  \xi_5  \ { 8 - (\kappa_{\omega}^{(0)})^8 (1+6 \, (\omega\bar\omega)^2 +
(\omega\bar\omega)^4)   \over (\omega\bar\omega+1)} \nonumber \\
&& \nonumber \\
&  = &   - \ 8 \ \xi_5  \ {(\omega\bar\omega-1)^4
 \over (\omega\bar\omega+1)^5}  \ ,
\label{4L_planar}
\ea
where $\xi_5$ is a numerical constant proportional to $\zeta(5)$ (see also \cite{PSZNoGo}),
and to obtain the second equality we used the order $g^2$ condition of
eq.~(\ref{1L_planar}).

It follows that even in the planar limit, in order to make the
theory finite we are obliged to fine-tune order by order in $g^2$
the coefficients in the superpotential. Indeed, for generic $\omega$ the above
equation necessarily requires a nonvanishing correction to the
superpotential.
The only exception is when $|\omega|=1$, in which case  the order $g^2$  condition alone
is sufficient for all order finiteness~\cite{PSZ2}.

Let us stress that eq.~(\ref{4L_planar}) is scheme independent, since
it follows from requiring the cancellation of the leading logarithmic
divergencies. However, as it
involves  two different (super)diagrams, the vanishing of the
divergent part does not automatically imply the cancellation of the
(possibly scheme dependent) finite parts.
These finite corrections change only the normalization of the
chiral superfield propagator and
will give rise to logarithmic divergencies at the next
order. Hence, even in the planar limit, starting from order $g^{10}$, the
explicit form of the condition for finiteness of the chiral
propagator will be scheme dependent.

Let us note also that
the 5-loop discrepancy in the planar Maximally Helicity Violating
(MHV) amplitudes pointed out in~\cite{Khoze}, is actually  
not present thanks to the order $g^8$ finiteness
condition in eq.~(\ref{4L_planar}). Indeed the vacuum diagram
considered in~\cite{Khoze} is equivalent to the 4-loop planar
correction to the chiral propagator shown in Figure~4,  which, as we have shown,
can be cancelled by fine-tuning the parameters in the superpotential.

To conclude, let us briefly comment on the freedom which the
conditions for finiteness of the chiral propagator leave. As we
already noted, all the coefficients $\kappa_{\Omega}^{(n)}$  remain
arbitrary. Thus one simple choice is $\kappa_{\Omega}^{(n)}=0$ for
all $n$. In this case the superpotential in
eq.~(\ref{superp}) reduces to the simpler expression~(\ref{old_superp}),
widely studied in the literature. However, as
noted in~\cite{PSZ3} (see also eq.~(\ref{2L_Omega})), even if we start with the simple
superpotential proportional only to $[ \ , \ ]_{\omega}$, the
quantum corrections to the  effective superpotential ${\cal
W}_{\rm eff}$ give rise also to contributions proportional to  $[
\ , \ ]_{\Omega}$. This suggests that
we may choose $\kappa_{\Omega}^{(n)}$ to precisely cancel the
quantum corrections to the
effective superpotential proportional to $[ \ , \ ]_{\Omega}$ ,  obtaining
\be
{\cal W}_{\rm eff} \sim   \ \tr \left( \Phi^{1} \ [\Phi^{2},
\Phi^{3}]_{\omega} \right) \ ,
\label{Weff}
\ee
to all orders in perturbation theory. For the first two coefficients one finds
in this case 
$\kappa_{\Omega}^{(1)}=0$ and
$\kappa_{\Omega}^{(2)}=-{\cal W}_{\rm eff} |_{g^5\, , \, \Omega} /N^2$,
where ${\cal W}_{\rm eff} |_{g^5\, , \, \Omega}$ is given in eq.~(\ref{2L_Omega}).

\section{The origin of the protected operator $\cO_F$}
\label{Free}

In this section we show that among all CPOs made out of matter superfields
the protected operator $\cO_F$~(\ref{defFr}) occupies a special place.
It can be derived directly from the Lagrangian~(\ref{1.1}) by
the so-called insertion procedure, i.e.\ by exploiting the information
that can be obtained in a superconformal theory by taking derivatives of
correlation functions with respect to the independent coupling constants.
Each such derivative gives rise to the insertion of a CPO which is a
combination of the SYM Lagrangian $\tr(W^2)$ and of terms from the
superpotential. In this context the protected operator
$\cO_F$ arises as the particular linear
combination of these CPOs which does not contain $\tr(W^2)$.
We show that the form of $\cO_F$ is directly determined from the
finiteness condition $\g_\P=0$. We confirm this result by an
explicit three-loop calculation.
The generalization to a superpotential with cubic terms
$\sum_I(\P^I)^3$ is straightforward and gives rise to a family of protected operators of this type.

\subsection{Quantum corrections through insertions}\label{proce}

Here we briefly describe a procedure which provides useful information
about the quantum corrections to Euclidean $n$--point correlation function
(for details see~\cite{Collins:1984xc}, Chapter 6.7).
Consider the expectation value
\begin{equation}\label{2.9}
  G \equiv \lan \cO(1)\cdots \cO(n)\ran =
\int e^{-\int d^4x_0d^4\q_0 \cL(x_0,\q_0,\bq_0;\, g_i)} \cO(1)\cdots \cO(n)
\end{equation}
where it is assumed that the Lagrangian depends of a set of independent coupling
constants, $g_i$. In order to avoid irrelevant (for the present discussion) contact
terms we will always take the operators  $\cO$ at unequal space-time points.

The quantum corrections to  the correlator~(\ref{2.9})
can be obtained by  differentiating  it with respect to the couplings $g_i$.
Each such derivative leads to the
insertion of a derivative of the action into the correlator~\footnote{The functional
integral in~(\ref{2.9}) should be divided by
$\int e^{-\int  \cL}$. Connected correlators are automatically generated in this way by
differentiation. To simplify notations we will not explicitly indicate this.}
\begin{eqnarray}
  \frac{\pa G}{\pa  g_i} &=& -\int e^{-\int \cL} \left[\int d^4x_0d^4\q_0\
   \frac{\pa \cL(x_0,\q_0,\bq_0;\, g_j)}{\pa  g_i} \right] \cO(1)\cdots \cO(n)\nonumber\\
  &=&  -\int d^4x_0d^4\q_0\  \lan  \frac{\pa\cL(0)}{\pa  g_i}
\cO(1)\cdots \cO(n)\ran \,, \label{2.10}
\end{eqnarray}
In what follows we assume that the theory is (super)conformal, i.e. all the beta functions
$\b_{g_i}$ vanish. As we already know, in the deformed theory this is achieved by imposing
a constraint on the couplings which should be taken into account when differentiating. We shall come back to this
essential point in Section \ref{CPOder}.

Before discussing the superconformal insertion procedure, let us explain some
details about its conformal analog. To be more specific, let us consider the
simplest case of scalar operators $\cO$ with  $n=2$. After integration over $d^4\q_0$ in~(\ref{2.10}) and
setting $\q=\bq=0$, we obtain
\begin{equation}\label{2.101}
  \frac{\pa }{\pa   g_i} \lan \cO(1) \cO^\dagger(2) \ran = -\int d^4x_0 \
  \lan  \frac{\pa L(x_0; g_j)}{\pa   g_i}   \cO(1)\cO^\dagger(2)\ran \,,
\end{equation}
where $L(x_0)$ is the Lagrangian operator (the top component in the $\q$ expansion of $\cL$).
The bare operators $\cO$ in~(\ref{2.101}) are
in general ill defined. Consequently, they must be renormalized,
unless they are protected. Let us start with the case of a single multiplicatively renormalized
operator. We will use the notation $[\cO](x) = \lim_{\ep\to 0} \hat\cO(x,\ep)$ with
$\hat\cO(x,\ep) = Z(\ep,\mu, g_i)\cO(x,\ep)$. Here $Z(\ep,\mu, g_i)$ is a renormalization factor
depending on the couplings $g_i$, on the regulator $\ep$ (e.g., a four-vector
$\vec\ep$ in point-splitting regularization) and on the renormalization scale
(or subtraction point) $\mu$. $\cO(x,\ep)$ is the regularized version of the bare operator
(e.g., with the constituent fields put at distances $\vec\ep$).
Now, suppose that the {\it renormalized} operator $[\cO]$ is a conformal primary of
dimension $\Delta = \Delta_0 + \g(g_i)$, where $\Delta_0$ is the naive
and $\g$ the anomalous dimension. In the point-splitting scheme we have~\cite{Bianchi:2003eg}
\begin{equation}\label{2.1001}
  Z(\ep,\mu, g_i) = (\ep^2\mu^2)^{-\g(g_i)/2}\,.
\end{equation}
Repeating the differentiation~(\ref{2.101}), but this time with the renormalization
factors included, we find
\begin{eqnarray}
  \frac{\pa }{\pa   g_i} \lan \hat\cO(x_1, \ep_1) \hat\cO^\dagger(x_2, \ep_2) \ran
&=& -\frac{1}{2} \frac{\pa\gamma}{\pa g_i}\, \ln(\ep_1^2\ep_2^2\mu^4) \ \lan
   \hat\cO(x_1, \ep_1) \hat\cO^\dagger(x_2, \ep_2) \ran   \nonumber  \\
  &-& \int d^{4}x_0 \  \lan  L_{g_i}(x_0)   \hat\cO(x_1, \ep_1)
\hat\cO^\dagger(x_2, \ep_2)\ran \, ,\label{2.1002}
\end{eqnarray}
where for short we have set $L_{g_i} = \pa L/\pa g_i$. Since we are taking the derivative of
a finite quantity, the apparent logarithmic singularity in the first term of the
right-hand side of eq.~(\ref{2.1002}) has to be compensated by a similar singularity in the second
term~\footnote{Our discussion about this point is similar to that in~\cite{Basu:2004nt},
except for the regularization scheme used.}.

To show how this comes about we recall that two-point correlator
of renormalized scalar operators takes the form predicted
by conformal invariance:
\begin{equation}\label{2.102}
 \lan [\cO](x_1)  [\cO]^\dagger(x_2) \ran = \lim_{\vec\ep_{1,2} \to 0}
\lan \hat\cO(x_1, \ep_1) \hat\cO^\dagger(x_2, \ep_2) \ran = \frac{A(g)}{(x^2_{12})^\Delta}\,,
\end{equation}
where $A(g)$ is a (coupling-dependant) normalization constant.
We remark that if $\Delta$ is an integer $\geq 2$, then the
distribution $1/x^{2\Delta}$ is singular, with a $\delta$- (or derivatives of $\delta$-)
function type singularity. Such contact terms become important
in the $n+1$--point correlators with the insertion~(\ref{2.10}), see below.

The derivatives of the Lagrangian $L$ with respect to the couplings must have the right conformal weight
in order to make the integral in~(\ref{2.1002}) conformally invariant (in the limit  $\vec\ep_{1,2}\to 0$).
In other words, we assume that the operators $L_{g_i} = \pa L/\pa g_i$ are conformal primaries
of dimension four. This assumption can be justified in a superconformal theory such as $\cN=4$ SYM
or its deformed $\cN=1$ version (see subsection~\ref{prim}).

Now, conformal invariance can also predict the form of the ``regular'' part of the
$2+1$--point correlator in the last term in~(\ref{2.1002}), yielding
\begin{equation}\label{2.104}
  \lan  L_{g_i}(x_0)   [\cO](x_1) [\cO]^\dagger(x_2) \ran_{\rm regular} =
  \frac{B_i(g)}{(x^2_{01} x^2_{02})^{2}(x^2_{12})^{\Delta-2}}\,.
\end{equation}
where ``regular'' means $x_0\neq x_1 \neq x_2$. Inserting~(\ref{2.104}) into~(\ref{2.1002})
leads to a divergent integral which we regularize by splitting points 1 and 2
\begin{equation}
 \frac{B_i(g)}{(x^2_{12})^{\Delta-2}}\int \ \frac{d^{4} x_0}{x^2_{01}x^2_{01+\ep_1}x^2_{02}x^2_{02+\ep_2}} =
  - \pi^2 \ln(\ep^2_1\ep^2_2\mu^4)\frac{B_i(g)}{(x^2_{12})^\Delta} + \mbox{finite part}\,, \label{2.106}
\end{equation}
where we have introduced the subtraction point, $\mu$. We now see that this term can provide
the singularity which cancels the logarithm in the first term in the
right-hand side\ of~(\ref{2.1002}), if
\begin{equation}\label{2.108}
  \frac{\pa\gamma}{\pa g_i} = 2\pi^2 \frac{B_i(g)}{A(g)}\,.
\end{equation}

The conclusion is that the anomalous dimension $\g(g)$ is controlled by the
regular part of the $2+1$--point correlator~(\ref{2.104}).
This observation, generalized to the supersymmetric case, explains why CPOs have
no anomalous dimension. The reason is that in both the deformed and undeformed
theories superconformal invariance forbids the existence of a nonvanishing 
$\lan  L_{g_i}(0)[\cO](1)[\cO]^\dagger(2)\ran_{ \rm regular}$
with $[\cO]$ being a CPO.
Consequently, such operators keep their naive dimension $\Delta_0$, i.e.,
they are ``protected". However, this does not necessarily mean the total absence
of quantum corrections to the correlator $\lan [\cO] [\cO]^\dagger \ran$.
Indeed, conformal invariance allows contact term contributions to the $2+1$--point correlator
of the form
\begin{equation}\label{2.105}
  \lan  L_{g_i}(x_0)   [\cO](x_1) [\cO]^\dagger(x_2) \ran_{ \rm contact} =
  C_i(g)\,\left[\delta^4(x_{01}) + \delta^4(x_{02}) \right] \frac{1}{(x^2_{12})^{\Delta_0}} \,.
\end{equation}
The appearance of such terms is related to the general fact that the
factors $1/x^4$ in~(\ref{2.104}) are singular distributions with a $\delta$-function
type singularity~\footnote{Note that in the case of CPOs with $\Delta_0$ an integer $\geq 2$,
an ultralocal contact term, like $\delta^4(x_{01})\delta^4(x_{02})$ for $\Delta_0=2$,
could be added to~(\ref{2.105}). We need not consider such terms here since
we are assuming $x_{12} \neq 0$.}.
It is clear that such terms, integrated over the insertion point $x_0$, will give quantum
corrections to the normalization $A(g)$ of the correlator~(\ref{2.102}).
This is precisely what happens to CPOs in the deformed theory, starting at
two loops~\cite{PSZ,Nie}. In the undeformed theory the more
powerful extended superconformal symmetry forbids even the contact terms, so there the
two-point functions of CPOs are completely protected (for more details see~\cite{HHS}).

We note that the above insertion procedure can be generalized in an obvious way
in the presence of mixing, i.e.\ when the renormalized operators
have the form $\hat\cO_i = Z_{ij}\cO_j$.

\subsection{CPOs as derivatives of the deformed $\cN=4$ Lagrangian}\label{CPOder}

\subsubsection{Holomorphic form of the action. Finiteness condition}

For our purposes in this section it is more convenient to rewrite the action of the deformed $\cN=4$ theory as follows:
\begin{eqnarray}
    S & = &
         \tr\left\{ \int  d^4x d^4\q  \,
     e^{-{gV}} \dP_{I}
    e^{{gV}}\Phi^{I} \right.    \nonumber \\
    & &  +\int d^4x_L  d^{2}\theta \,  W^{\alpha}W_{\alpha}  \nonumber \\
    &  & +\,   \!\!  \int
d^4x_L d^{2}\theta \, \left(
     f\,\Phi^{1}[\Phi^{2},\Phi^{3}] + d\,\Phi^{1}\{\Phi^{2},\Phi^{3}\} \right) \nonumber \\
     && + \,  \left. \int
d^4x_R d^{2}{\bar
    \theta} \, \left( \bar f\,
    \dP_{1}[\dP_{3},\dP_{2}] +  \bar d\,
    \dP_{1}\{\dP_{3},\dP_{2}\} \right)
    \rule{0pt}{18pt}   \right\} \; ,
    \label{1.1}
\end{eqnarray}
where $W_{\alpha}$ is the chiral field-strength of the gauge potential
$V$,
\be\label{1.2}
W_{\alpha}= -{1 \over 4g} \bar {D^{2}} \left( e^{-{2gV}} D_{\alpha}
 e^{{2gV}} \right) = -{1 \over 2} \bar {D^{2}} D_{\alpha} V + O(g) \,.
\ee
Note that in~(\ref{1.1}) we have written the SYM kinetic term in its chiral form.
If needed, it can be rewritten as an antichiral term according to the
identity $\int d^{2}\theta \:  W^2 = \int  d^{2}\bq \:  \bar W^2$ valid in a
topologically trivial background.\footnote{It is more convenient, although
not essential, to work with a real gauge coupling $g$, therefore we do not introduce an
instanton angle.} Unlike Section \ref{sec:Allloops}, here we prefer to parametrize
the potential by two complex coupling
constants, $f$ in front of the commutator (i.e., color tensor $f_{abc}$) and $d$ in
front of the anticommutator (i.e., color tensor $d_{abc}$). Both of them as well as the
gauge coupling $g$ are treated as small perturbative parameters $f\sim d\sim g$.
To switch back to the notation of Section \ref{sec:Allloops} we need to perform the change of parametrization
\begin{eqnarray}
  f &=& \frac{g}{{2}}\,[(\o+1)\kappa_\o(g) + (\Omega+1)\kappa_\Omega(g)]   \nonumber\\
   d &=& \frac{g}{{2}}\,[(\o-1)\kappa_\o(g) + (\Omega-1)\kappa_\Omega(g)] \ . \label{r}
\end{eqnarray}

The deformed theory is finite (and hence conformal) up to three loops~\footnote{We remark
that now the counting of ``loops" is somewhat different than in the preceding section.
There all the couplings were expressed in terms of $g$, so ``$n$ loops" meant perturbative
level $g^{2n}$. Here ``$n$-loop" corrections are homogeneous polynomials of degree $2n$
in the couplings $g,f,d$.} if the couplings satisfy the relation
\begin{equation}\label{1.6}
   \gamma(g,{f,d}, \bar f, \bar d)= \g^{(1)} +  \g^{(3)} = 0 \, .
\end{equation}
This is the condition for vanishing anomalous dimension $\g_\Phi$ of the matter superfields $\P$ involving a one-loop
\begin{equation}\label{1.6'}
    16 \pi^2 \g^{(1)} =    2\frac{N^2-4}{N} |d|^2 + 2N |f|^2 -2N g^2
\end{equation}
and a three-loop
\begin{equation}\label{1.6''}
    (16 \pi^2)^3 \g^{(3)} = - \frac{ 24\zeta(3) ({N^2-4}) }{{N^3}}  |d|^2
     \big[ 2({N^2-10}) |d|^4 -3  {N^2}  ({f^2}\bar d^2 +{d^2}  \bar{f}^2)\big]
\end{equation}
contributions \footnote{This condition is a particular case of the finiteness condition
for the most general trilinear superpotential~(\ref{gensupepo}) discussed in~\cite{JackJN}.
It was obtained in its explicit form for the case of  the deformed $\cN=4$ SYM potential
in~\cite{FG} at one loop, in~\cite{PSZ} at two loops and in~\cite{Nie} at three loops.  }
(cf. eq.~(\ref{3loopv1})). The absence of a two-loop contribution is explained by the fact that in carrying out
the two- and three-loop calculation of $\g_\Phi$ the one-loop condition $\g^{(1)} = 0$
has been used, after which no new condition arises at two loops. Alternatively,
the calculation may be done without imposing the one-loop condition.
In this case one finds~\cite{JackJN} a two-loop contribution to $\g_\Phi$ of the form
\begin{equation}\label{1.6'''}
    \g^{(2)} = \g^{(1)}\ P^{(1)}
\end{equation}
where $P^{(1)}$ is some homogeneous polynomial of degree two in the couplings whose explicit
form is not important for us. Similarly, $\g^{(3)}$ gets modified by a term of the
type $\g^{(1)}\ P^{(2)}$ with $P^{(2)}$ a polynomial of degree four. Thus, the complete
version of the finiteness condition~(\ref{1.6}) up to three loops has the form
\begin{equation}\label{comp3}
    \g^{(1)}(1+P^{(1)} + P^{(2)}) +  \g^{(3)} = 0 \, .
\end{equation}
Clearly, since $P^{(1)}, \ P^{(2)} << 1$,  we can multiply this equation by $(1+P^{(1)} + P^{(2)})^{-1}$.
In this way we recover (\ref{1.6}), up to terms of four-loop order which are beyond the scope of this section.

Following~\cite{JackJN}, we note that the three-loop condition $\g^{(1)} +  \g^{(3)} = 0 $
can be formally reduced to a one-loop condition of the type $\g^{(1)} = 0$ by a change of
variables (or, equivalently, by a finite coupling renormalization). For instance, one such change is
\begin{equation}\label{chvarcond}
    f \ \rightarrow \ f - \frac{9 \zeta(3) (N^2-4)}{64 \pi^4 N^2}\, d^3 \bar d \bar f\,,
     \qquad d \ \rightarrow \ d + \frac{3 \zeta(3) (N^2-10)}{64 \pi^4 N^2}\, d^3 \bar d^2\,.
\end{equation}

Finally, returning to the notation of Section \ref{sec:Allloops}, it is easy to see that the
condition~(\ref{1.6}) with $\g^{(1)}$ from~(\ref{1.6'}) and $\g^{(3)}$ from~(\ref{1.6''}),
rewritten in terms of $\kappa_{\o,\Omega}(g)$ as indicated in~(\ref{r}) and expanded up to $g^6$,
is equivalent to~(\ref{1Lcond}) at order $g^2$,~(\ref{2Lcond}) at order $g^4$ and~(\ref{3loopv1})
at order $g^6$.

\subsubsection{Holomorphic derivatives. CPOs as generators of quantum corrections}\label{holde}

The general procedure which generates quantum corrections to the $n$-point correlation
functions~(\ref{2.9}) was described in section~\ref{proce}.  The derivative with respect
to each independent coupling gives rise to an operator insertion into the correlator,
as shown in~(\ref{2.10}). Below we prove that in the specific case of the
action~(\ref{1.1}) this procedure amounts to the insertion of {\it chiral or antichiral primary operators}.

The condition for finiteness~(\ref{1.6}) means that in our case the couplings, and hence their
 variations  are not independent, rather they satisfy, 
\begin{equation}\label{002}
  \g_g \d g + \g_f \d f + \g_d \d d + \g_{\bar f}\d \bar f + \g_{\bar d}\d \bar d = 0\,,
\end{equation}
which implies for the variation of the $n$-point correlator $G$
\begin{eqnarray}
  \d G&=&G_g \d g + G_f \d f + G_d \d d + G_{\bar f}\d\bar f + G_{\bar d} \d\bar d \nonumber\\
  &=&\left(G_f - \frac{\g_f}{\g_g}\, G_g  \right) \d f + \left(G_d - \frac{\g_d}{\g_g}\, G_g
  \right) \d d + \mbox{c.c.} \ . \label{001}
\end{eqnarray}
In this equation we have treated the holomorphic couplings $f$ and $d$ as independent,
while the gauge coupling is taken as a (real) function of them, $g=g(f,d,\bar f,\bar d)$. This point of
view is preferable here because we want to obtain {\it chiral} operators through differentiation
with respect to the {\it holomorphic} couplings $f,d$. In Section \ref{sec:Allloops} we adopted
the alternative (i.e.\ {\it perturbative}) point of view where the gauge coupling is the
universal small parameter used in the perturbative calculations.

In order to compute the derivatives of the Lagrangian in~(\ref{1.1}) (completed with  the
gauge-fixing term $\xi\int  d^4x d^4\q  \, D^2V \bar D^2V $ and with the ghost term) with
respect to the independent holomorphic couplings, we first absorb the gauge coupling $g$ into
the gauge potential and the gauge-fixing parameter\footnote{In~\cite{Arkani-Hamed:1997mj}
it is claimed that this redefinition of $V$ may
lead to the so-called ``rescaling anomaly", which is used to justify the ``exact" NSVZ beta
function~\cite{Novikov:1983uc}. However, as mentioned in~\cite{Arkani-Hamed:1997mj},
the rescaling anomaly is not seen
in dimensional regularization. Here we adopt the point of view that there exists a
scheme free from such anomalies. Note also that the rescaling of the gauge-fixing
parameter $\xi$ in~(\ref{2.11}) has no effect on the correlators~(\ref{2.9})
of {\it gauge invariant} composite operators $\cO$.}
\begin{equation}\label{2.11}
  V \ \rightarrow \ g^{-1}V\, , \qquad \xi \ \rightarrow \ g^2\xi\, .
\end{equation}
The effect of this rescaling is that $g$ now appears only in front of the classical
SYM term $W^2$ in the Lagrangian (recall~(\ref{1.2})), while it drops out from the
gauge-fixing and ghost terms. So 
\begin{equation}\label{003}
  \cL_g = -\frac{2}{g}\,\tr ( W^2 ) \,,
\end{equation}
while the variation with respect to the holomorphic couplings gives
\begin{equation}\label{004}
    \cL_f = \tr \left(\Phi^{1}[\Phi^{2},\Phi^{3}] \right) \,,
    \qquad \cL_d = \tr \left(\Phi^{1}\{\Phi^{2},\Phi^{3}\} \right)\,.
\end{equation}

It is now clear that the total derivative of $G$ with respect to each {\it holomorphic}
coupling gives rise to the insertion of a {\it chiral operator}:
\begin{eqnarray}
  \frac{\d G}{\d f}&=&  -\int\, d^4x_L d^2\q_0\  \lan  \cI_f(0)  \cO(1)\cdots \cO(n)\ran\nonumber\\
  \frac{\d G}{\d d}&=&  -\int\, d^4x_L d^2\q_0\  \lan  \cI_d(0)  \cO(1)\cdots \cO(n)\ran \label{005}
\end{eqnarray}
where
\begin{eqnarray}
  \cI_f&=& \frac{2\g_f}{g\g_g}\,\tr (W^2) +  \tr\left(\Phi^{1}[\Phi^{2},\Phi^{3}]\right) \nonumber\\
  \cI_d&=& \frac{2\g_d}{g\g_g}\, \tr (W^2) + \tr\left(\Phi^{1}\{\Phi^{2},\Phi^{3}\}\right) \,. \label{006}
\end{eqnarray}
In section \ref{prim} we show that the insertions $\cI$ are chiral {\it primary}
operators of protected conformal dimension $\Delta_0=3$. Similarly, differentiating with respect
to the antiholomorphic couplings amounts to inserting antichiral operators (to this end the
 SYM kinetic term should be rewritten in its antichiral form).

This result admits the following interpretation\footnote{ES is grateful to Ken
Intriligator for this remark.}. The finiteness condition~(\ref{1.6}) can
be viewed as the equation of a real hypersurface in the complex moduli space
 of the couplings.
In Section \ref{sec:Allloops} we chose to describe it in a parametric fashion, by expressing the matter
couplings as functions (perturbative power series) of the gauge coupling $g$. In this section
we use the holomorphic couplings $f$ and $d$ as the independent coordinates on the surface whose
tangent space is spanned by the derivatives $\pa/\pa f$ and $\pa/\pa d$. The quantum
equivalents of the tangent space vectors are the CPOs~(\ref{006}). Each of them generates quantum
corrections to the Green's functions of operators in the theory when moving along the corresponding
tangent direction to the surface of couplings.

Finally, the operator $\cO_F$~(\ref{defFr}) is nothing but the linear combination of $\cI_f$ and $\cI_d$
from which the SYM term $W^2$ drops out:
\begin{equation}
  \cO_F = \frac{\g_d \cI_f - \g_f \cI_d}{\g_d - \g_f} = \tr\left( \Phi^{1}\Phi^{2}\Phi^{3} \right)
  + \frac{\g_f +
  \g_d}{\g_f -\g_d} \tr\left( \Phi^{1}\Phi^{3}\Phi^{2} \right)\,.
     \label{007}
\end{equation}
The explicit form of the relative coefficient in~(\ref{007}) is determined by a straightforward
calculation. From~(\ref{1.6}),~(\ref{1.6'}) and~(\ref{1.6''}) we find
\begin{eqnarray}
   \frac{\g_f + \g_d}{\g_f -\g_d} &=& \frac{N^2 \bar f   +(N^2-4) \bar d}{N^2 \bar f -  (N^2-4) \bar d}
    \ \ - \  \ \frac{9\, \zeta(3)\big({N^2-4}\big)}{32 \pi^4}\ \bar d\  \times   \label{007'} \\
  &&   \nonumber \\
   &&  \frac{2 ({N^2-4}){{\bar d}^3}  d  f  -3   {N^2} {{\bar f}^3}  {d^2}
     -  {N^2} {{\bar d}^2}  \bar f  {f^2}+2 ({N^2}-10) {d^2} {{\bar d}^2}
    \bar f}{{\big[N^2 \bar f - (N^2-4) \bar d\big]^2}} \, . \nonumber
\end{eqnarray}
The first term coincides with the one-loop result first obtained in~\cite{FG}. The second term
is the three-loop correction; it has been obtained by expanding $(\g_f + \g_d)/(\g_f -\g_d)$
in $f \sim d \sim g$ up to $g^4$. The absence of a two-loop correction in~(\ref{007'}) (independently
noticed in~\cite{PSZ3}) is due to the specific form~(\ref{1.6'''}) of the two-loop contribution  which
allows us to rewrite the finiteness condition up to three loops in the form~(\ref{1.6}).

\subsubsection{Perturbative calculation at order $g^6$}

We have checked by an explicit computation that the operator
defined in eqs.~(\ref{007}),~(\ref{007'}) has vanishing anomalous dimension at order $g^6$.
Note that  with the notation introduced in eqs.~(\ref{defcom}),~(\ref{Omega}),
we can rewrite $\cO_F$ (up to an irrelevant rescaling) in the form
\be
\cO_F \sim  \tr\left(\Phi^{1} \ [\Phi^{2},
\Phi^{3}]_{\Omega} \right) + ( a_2 g^2 + a_4  g^4  + \dots) \ \tr\left( \Phi^{1} \ [\Phi^{2},
\Phi^{3}]_{\omega} \right)    \, .
\label{Friedman}
\ee
We have computed through order $g^6$ the corrections
to the two-point functions of (the lowest $\theta$ component of)
the operator  $\cO_F$ with all the operators it can mix with.
There are only three such operators, namely
\be
\tr\left(\Phi^{\dagger}_{1} \ [\Phi^{\dagger}_{3}, \Phi^{\dagger}_{2}]_{\bar \Omega} \right) \ , \qquad
\tr\left( \Phi^{\dagger}_{1} \ [\Phi^{\dagger}_{3}, \Phi^{\dagger}_{2}]_{\bar \omega} \right) \quad
{\rm and} \qquad \tr( \bar W^2) \ ,
\label{FriedmanMix}
\ee
All the corrections to the two-point function of $\cO_F$ with $\tr( \bar W^2)$,
as well as the logarithmically divergent correction to the two-point function
with $\tr (\Phi^{\dagger}_{1} \ [\Phi^{\dagger}_{3}, \Phi^{\dagger}_{2}]_{\bar \Omega} )$
vanish by the color contractions. At order $g^6$ the logarithmically
divergent correction to the two-point function  with
$\tr( \Phi^{\dagger}_{1} \ [\Phi^{\dagger}_{3}, \Phi^{\dagger}_{2}]_{\bar \omega} )$
comes only from the three superdiagrams depicted in Figure~5. The first two are genuine
order $g^6$ diagrams, while the last (order $g^2$) diagram appears multiplied
by the  order $g^4$ coefficient $a_4$ in~(\ref{Friedman}).
All three diagrams lead to the same (logarithmically divergent) coordinate space integral.
The cancellation of this divergence, i.e. the vanishing of the order $g^6$ anomalous dimension
of $\cO_F$, fixes the values of $a_2$ and $a_4$. Exactly the same values are obtained from
~(\ref{007'}) by rewriting it in the notation of Section \ref{sec:Allloops} (see~(\ref{r}))
and expanding in $g$ up to $g^4$.

\begin{figure} [ht]
    \centering
    \includegraphics[width=1\linewidth]{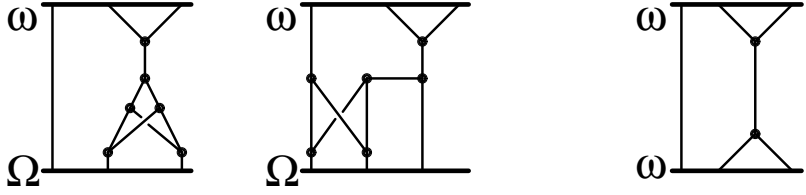}
       \vskip -200pt
\caption{Diagrams contributing to the 2-point functions of $\cO_F$
at order $g^6$.}
\end{figure}

\subsubsection{General potential with a single finiteness condition}\label{genpotsing}

The chiral superpotential in eq.~(\ref{1.1}) has the key feature that it leads to a single finiteness
condition. The reason for this is that the matrix of $\g$ functions of the matter superfields $\P^I$
is proportional to the unit matrix in flavor space,
\begin{equation}\label{umat}
    (\g_\P)^I_J = \g\, \delta^I_J\,,
\end{equation}
so it is enough to demand $\g = 0$ to ensure finiteness for all fields. In~\cite{LS} a generalized
superpotential with such a property was proposed in the form
\begin{equation}\label{z3}
    f\, \tr\left(\P^1[\P^2,\P^3]\right) + d\, \tr\left(\P^1\{\P^2,\P^3\}\right) +
    k\,\tr\left( (\P^1)^3 + (\P^2)^3 + (\P^3)^3\right)\,.
\end{equation}
Here the holomorphic couplings $f,d,k$ are subject to a single real condition generalizing~(\ref{1.6}):
\begin{equation}\label{gencond}
    \gamma(g,f,d,k, \bar f, \bar d, \bar k) = 0\,.
\end{equation}
Note that for generic values of the couplings the $U(1)\times U(1)$ symmetry of  the superpotential
is broken, only $U(1)_R$ survives (in addition, there is a discrete symmetry $Z_3 \times Z_3$).
It can be shown~\cite{Aharony:2002hx} that~(\ref{z3}) is in fact  the most general superpotential
of this type, up to field redefinitions in the form of $SU(3)$ transformations (recall that the
 matter kinetic term $\tr(\P^\dag_I \P^I)$ is $SU(3)$ invariant).

Just as in section~\ref{holde}, the derivatives of the Lagrangian with respect
to each independent holomorphic coupling (we treat $g$ as dependent
and $f,d,k$ as independent) lead to insertion formulae of the
type~(\ref{005}). This means that the quantum corrections are now generated by the three CPOs
\begin{eqnarray}
  \cI_f&=& \frac{2\g_{f}}{g\g_g} \tr(W^2) + \tr\left( \P^1[\P^2,\P^3]\right)\nonumber\\
  \cI_d&=& \frac{2\g_{d}}{g\g_g} \tr(W^2) +  \tr\left(\P^1\{\P^2,\P^3\}\right)\nonumber\\
  \cI_k&=& \frac{2\g_{k}}{g\g_g} \tr(W^2) +  \tr\left((\P^1)^3 + (\P^2)^3 + (\P^3)^3 \right)\,. \label{3gene}
\end{eqnarray}
From them we can form a one-parameter family (up to an overall factor)
of protected operators analogous to $\cO_F$ in eq.~(\ref{007})
\begin{equation}\label{famfri}
    h_1 \, \tr\left(\P^1[\P^2,\P^3]\right) + h_2 \, \tr\left(\P^1\{\P^2,\P^3\}\right)
     + h_3 \, \tr\left((\P^1)^3 + (\P^2)^3 + (\P^3)^3\right)
\end{equation}
with the coefficients satisfying the relation
\begin{equation}\label{relco}
    h_1\g_{f} + h_2\g_{d} + h_3\g_{k} = 0\,.
\end{equation}
Using the one-loop finiteness condition from~\cite{PSZ3}, we immediately
reproduce their form of the dimension three  protected operator.

\subsection{Operator mixing. Superconformal primaries and descendants}\label{prim}

In this subsection we justify our claim that the {\it chiral} insertions~(\ref{006}) are
also superconformal {\it primaries} and hence
they are protected operators (or CPOs). We do this by examining the mixing of all the
chiral terms in the Lagrangian (SYM kinetic term and matter superpotential).

Whether an operator is primary or not is a subtle question which can
only be answered at the quantum level. In~\cite{Witten:1998qj} a ``rule of thumb"
for CPOs made out of matter superfields was proposed, which says that ``an operator
is primary if it does not contain commutators of superfields under a single color
trace''. The presence of a commutator is, in fact, a signal that
the operator has been obtained from another, lower dimension operator via the field
equations, e.g., $\bar{\cal D}^2\dP \sim g[\P,\P]$ in the undeformed theory. However,
we know that this rule works only in the simplest cases. A counterexample are the
$1/4$ BPS operators~\cite{Bianchi:1999ge} which are mixtures
of single and double trace operators, the former containing commutators. This is a typical
case of {\it operator mixing}.

\subsubsection{Deformation with $f,d$ terms}

In a quantum theory  operators can mix if they have the same  quantum numbers.
For instance, the chiral terms in the Lagrangian~(\ref{1.1})
\begin{equation}\label{2.24}
  F =  \tr(W^2)\,, \qquad B_1 = \tr\left(\Phi^{1}[\Phi^{2},\Phi^{3}]\right) \,,
   \qquad B_2 = \tr\left(\Phi^{1}\{\Phi^{2},\Phi^{3}\} \right)
\end{equation}
are scalars of dimension 3 and of $R$ charge $2/3$ (in units in which $\q$ has charge $1$).
They also have vanishing $U(1)\times U(1)$ charges generated by the currents~\footnote{For
brevity the covariantizing factors
$e^{\pm 2gV}$ are suppressed.}
\begin{eqnarray}
  \cV_X &=& \tr\left( 2 \,  \dP_{1}
     \Phi^{1} -   \dP_{2}
     \Phi^{2}-  \dP_{3}
     \Phi^{3}\right) \nonumber\\
  \cV_Y &=& \tr\left( \dP_{2}
     \Phi^{2}-  \dP_{3}
     \Phi^{3}\right) \ .  \label{2.27}
\end{eqnarray}
The conservation of these currents
\begin{equation}\label{conscurre}
    D^2\cV_{X,Y} =\bar D^2\cV_{X,Y} =0\,,
\end{equation}
follows from the field equations of the Lagrangian~(\ref{1.1}).
The conclusion of this analysis is that the operators~(\ref{2.24}) can mix
among themselves.
The combinations $\cI_{f,d}$ in eq.~(\ref{006}) are two such mixtures.
Notice that $F = O(g^0)$ (recall~(\ref{1.2})), while $B_{1,2}$ appear
in~(\ref{006}) multiplied by the couplings $f\sim d\sim g$. This means
that the leading term in~(\ref{006}), i.e.\ the one that survives in the
free theory ($g=f=d=0$), is $F$, while the appearance of $B_{1,2}$ is a
quantum effect.

We can construct a third mixture of the same three operators $B_{1,2}$ and $F$,
where their roles are exchanged: $B_{1,2}$ are the leading terms and $F$ comes about
because of quantum corrections. This mixing pattern can also be seen as originating from
the so-called ``Konishi anomaly''~\cite{PiguetKonishi}. The Konishi operator is the sum of
the three matter kinetic terms in the Lagrangian~(\ref{1.1}),  $\cK = \tr( \dP_{I}\Phi^{I})$.
Hitting it with $\bD^2$ and using the {\it classical} field equations, we obtain
\begin{equation}\label{2.26}
  \bD^2 \cK = 3 (f\, B_1 + d\, B_2)\,.
\end{equation}
Unlike the currents~(\ref{2.27}) the Konishi operator is not conserved and hence not protected.

In the quantum theory the classical (non-conservation) equation~(\ref{2.26})
has to be corrected by an anomaly term. For example, at one loop one finds
\begin{equation}\label{2.28}
  \bD^2 \cK = 3 (f\, B_1 + d\, B_2) +  ag^2\, F \equiv g \cK'\,,
\end{equation}
where $a$ is some computable numerical coefficient~\footnote{In the literature there are
claims that the Konishi anomaly is one-loop exact (see, e.g.~\cite{Shifman:1999mv}).
Such claims fail to take into account the presence of the (classical) $B$ terms and
their non-trivial mixing at the quantum level with the operator $F$. This is most clearly
seen if one repeats the two-loop calculation of~\cite{Grisaru:1985ik}
{\it in the presence of a matter self-coupling} (ES thanks Marc Grisaru for discussions on
this point). See also~\cite{Eden:2005ve} for a general discussion of the Konishi
anomaly in the context of $\cN=4$ SYM.}. In fact, the quantum corrected equation
defines the superdescendant of the Konishi multiplet $\cK'$. Although not a
superconformal primary,  $\cK'$ is an operator with well-defined conformal
properties (conformal primary).

Returning to the first two combinations~(\ref{006}), we can now argue that they are
{\it superconformal primaries}. Indeed, if they were descendants, we should be able
to find some lower dimensional primaries from which we could obtain~(\ref{006})
through supersymmetry transformations (or, equivalently, through spinor derivatives,
as in~(\ref{2.28})). Since the leading $F$ terms in~(\ref{006}) are fermion bilinears,
we can only use two supersymmetries on a scalar operator of dimension two, or one
supersymmetry on a spinor of dimension 5/2. Given the $U(1)\times U(1)\times U(1)_R$
charges of $B_{1,2}$, the only scalar candidates are the three matter kinetic terms.
Two combinations are the $U(1)\times U(1)$ currents~(\ref{2.27}),
which have no descendants  obtained acting  with $\bD^2$. The third one is the
Konishi operator $\cK$ which gives rise to the descendant $\cK'$. Similar arguments
rule out a fermionic primary. Thus, the operators~(\ref{006}) must be primary, and since
they are also chiral, they are protected (or CPOs).

We remark that the protected operators are orthogonal to the
Konishi descendant (the latter has a non-vanishing anomalous dimension,
the former do not), implying
\begin{equation}\label{2.30'}
  \lan  \cK'\ \cI^\dagger_{f,d} \ran =  0\,.
\end{equation}
This condition can be efficiently used for determining the right mixture
in~(\ref{2.28}) not only at one loop, but also beyond
(see~\cite{I,Eden:2005ve})\footnote{In the recent paper~\cite{PSZ3} the logic has been
inverted, using the orthogonality of the protected operator $\cO_F$ to the Konishi
descendant as a tool for determining the former from the known form of the latter.
However, experience shows that the direct determination of the Konishi
anomaly beyond one loop is not an easy task.}.

\subsubsection{General deformation with $f,d,k$ terms}

In the case of the superpotential~(\ref{z3}) the $U(1)\times U(1)$ symmetry is broken and
only $U(1)_R$ survives. This means that we can extend the set of chiral
operators~(\ref{2.24}) by adding the new terms
\begin{equation}\label{addop}
    B_3 =\tr\left( (\Phi^{1})^3\right)
    \,, \qquad B_4 = \tr\left((\Phi^{2})^3\right)
    \,, \qquad B_5 = \tr\left((\Phi^{3})^3\right)
    \,.
\end{equation}
appearing in~(\ref{z3}). Out of the set of six operators~(\ref{2.24})
and~(\ref{addop}) we can form the three protected combinations of eq.~(\ref{3gene})
and three new, unprotected  combinations. One of the latter is, as before, the Konishi descendant
\begin{equation}\label{2.28'}
  \bD^2 \cK = 3 \left(f\, B_1 + d\, B_2 + 3 k (B_3+B_4+B_5)\right) +  ag^2\, F \equiv g \cK'\,.
\end{equation}
The two new ones are descendants of the former $U(1)\times U(1)$
currents~(\ref{2.27}) which are not conserved anymore
\begin{equation}\label{notcons}
    \bar D^2\cV_{X} = 3k (2 B_3 - B_4 - B_5)\,, \qquad \bar D^2\cV_{Y} = 3k (B_4 - B_5)\,.
\end{equation}
If we now switch off the new deformation by setting $k=0$, conservation of currents
is restored, while at the same time the chiral operators $2 B_3 - B_4 - B_5$ and
$B_4 - B_5$ become primary and thus protected.

\subsubsection{Undeformed theory}

In the undeformed $\cN=4$ theory we have $f=g$ and $d=k=0$ which restores
the full $SU(3)\times U(1)_R \subset SU(4)$ symmetry.
The Konishi operator has a scalar descendant $\cK_{10}$ of dimension three
in the \underline{10} of $SU(4)$.
Its $SU(3)$ singlet projection $\cK_{10/1}$~\footnote{We denote by $m/n$
the $n$-dimensional $SU(3)$ projection of an $m$--dimensional $SU(4)$ multiplet.}
in the $\cN=1$ formulation of the theory is the Konishi anomaly, a mixture
of $B_1$ and $F$, which is the analog of $\cK'$. The same $SU(3)$ invariant
operators $B_1$ and $F$ form another combination,
\begin{equation}\label{2.281}
  \cO_{10/1} = 2F-gB_1=-g\cI_f\,.
\end{equation}
We observe that $\cO_{10/1}$ is on the one hand proportional to $\cI_f$ of eq.~(\ref{006})
computed with $\g =2N (|f|^2 - g^2)=0$, $d=0$. On the other hand, it is a descendant
{\it in the $\cN=4$ sense} of the protected stress-tensor multiplet $\cO_{20'}$.
It can be obtained by making two on-shell $\cN=4$ supersymmetry
transformations on the highest-weight chiral projection $\cO_{20'/6} = \tr( \P^1\P^1)$
of the $1/2$ BPS operator $\cO_{20'}$,
\begin{equation}\label{2.282}
  \cO_{10/1} = (\bar Q_{\cN=4})^2 \cO_{20'/6}\,.
\end{equation}
It is important to realize that  $\cO_{10/1}$ is not a descendant
of $\cO_{20'}$ in the $\cN=1$ sense. This confirms that $\cI_f$
is a superconformal primary from the $\cN=1$ point of view.

Finally, in the undeformed case the operator $\cO_F$ eq.~(\ref{007})
is a particular state in the $SU(3)$ $10$-plet projection $\cO_{50/10}$
of the protected $1/2$ BPS operator $\cO_{50}$
whose highest weight is $\tr((\P^1)^3)$.

\section{Conclusions}

We have shown that for {\it any} complex value of the deformation parameter
$\omega$ there exists a whole family (parametrized by the  complex function
$\kappa_{\Omega}(g)$) of conformally invariant $\cN=1$ deformations of  ${\cal N}=4$ SYM.
In each such theory is present a special CPO  $\cO_F$, of dimension three,
whose explicit form to all orders can be determined directly from the finiteness condition
 $\g_\P=0$.

In the recent paper~\cite{PSZNoGo} the planar limit of the deformed
${\cal N} = 4$ SYM theory has been investigated in detail up to ten loops.
Our, order $g^8$ planar, calculation is in agreement
with the four-loop formula obtained there.
However, we disagree with the conclusions of~\cite{PSZNoGo},
where it is claimed that the deformed ${\cal N} = 4$ SYM can be made conformally
invariant only if the deformation parameter $\beta$ is real
(i.e. for $|\omega|=1$ in our notation). We believe that
the contradiction they find at the five-loop level is an artefact of
the use of dimensional regularization and that
the solution to this problem is given in~\cite{Ermushev:1986cu,Kazakov:1986vg}.

\section*{Acknowledgements }
 ES is grateful to G. Arutyunov, P. Heslop, P. Howe, K. Intriligator, D. Kazakov and M. Strassler
 for enlightening discussions. ES acknowledges the warm hospitality extended to
 him by Massimo Bianchi and the whole theory group at the University of Rome ``Tor Vergata"
 where part of this work has been carried out. This work was supported in
part by INFN, by the MIUR-COFIN contract 2003-023852, by the MIUR-PRIN contract 2004-, 
by the EU contracts MRTN-CT-2004-503369 and MRTN-CT-2004-512194, by the
INTAS contract 03-516346 and by the NATO grant PST.CLG.978785.

\end{document}